# Microwave neural processing and broadcasting with spintronic nano-oscillators

P. Talatchian[1], M. Romera[1], S. Tsunegi[2], F. Abreu Araujo[1,3], V. Cros[1], P. Bortolotti[1], J. Trastoy[1], K. Yakushiji[2], A. Fukushima[2], H. Kubota[2], S. Yuasa[2], M. Ernoult[1,4], D. Vodenicarevic[4], T. Hirtzlin[4], N. Locatelli[4], D. Querlioz[4], J. Grollier[1]

[1]Unité Mixte de Physique, CNRS, Thales, Univ. Paris-Sud, Université Paris-Saclay, France, email: julie.grollier@cnrs-thales.fr
[2]National Institute of Advanced Industrial Science and Technology (AIST), Spintronics Research Center, Japan
[3]Institute of Condensed Matter and Nanosciences, UCLouvain, Belgium
[4] Centre de Nanosciences et de Nanotechnologies, CNRS, Univ. Paris-Sud, Université Paris-Saclay, France

*Abstract—* Can we build small neuromorphic chips capable of training deep networks with billions of parameters? This challenge requires hardware neurons and synapses with nanometric dimensions, which can be individually tuned, and densely connected. While nanosynaptic devices have been pursued actively in recent years, much less has been done on nanoscale artificial neurons. In this paper, we show that spintronic nano-oscillators are promising to implement analog hardware neurons that can be densely interconnected through electromagnetic signals. We show how spintronic oscillators maps the requirements of artificial neurons. We then show experimentally how an ensemble of four coupled oscillators can learn to classify all twelve American vowels, realizing the most complicated tasks performed by nanoscale neurons.

## I. Spintronic nano-oscillators

Spintronic nano-oscillators are magnetic tunnel junctions, whose CMOS-compatible technology is essentially identical to magnetic memory cells that can be fabricated by billions on a chip [1]. They are cylinder shaped, with a diameter that can be reduced below 10 nm. The central part is a tunnel barrier (usually MgO) separated by two ferromagnetic layers (often based on cobalt, iron and boron alloys). As illustrated in Fig. 1, when a direct current is sent through the cylinder, it gets spin-polarized by crossing the first magnetic layer, tunnels to the other one, to which it transfers its excess of angular momentum by exerting a torque on the magnetization [2]. This spin-torque can then induce sustained magnetization precessions which are converted through magnetoresistance to voltage oscillations that can reach up to a few millivolts across the junction (Fig.2) [3]. The power consumption of spin-torque nano-oscillators, which decreases with their diameter, is around one microwatt for a diameter of ten nanometers. The frequency of the oscillations varies from hundreds of megahertz to tens of gigahertz. While the power consumption per se is not weak weak compared to slower CMOS-based neurons, the energy can therefore be very small, below a hundred attojoules per oscillation [4]. This means that hundreds of millions of oscillators could be assembled on a chip and used for computing.

## II. Spin-torque nano-oscillators can compute a non-linearity and broadcast the result

### A. Spin-torque nano-oscillators are non-linear nano-radios

The key feature of spin-torque nano-oscillators that makes them interesting as neurons is their non-linear response. As can be observed in Fig. 3, when the dc current through the oscillator is varied, the angle with which magnetization precesses varies widely, resulting in a non-linear dependence of the oscillation voltage amplitude. This means that, just like formal neurons in deep networks, spin-torque nano-oscillators can non-linearly transform their input (here the injected current). Moreover, the output of the computation is a microwave voltage or magnetic field that can be emitted. In other words, spin-torque-nano-oscillators can non-linearly process information like neurons, then broadcast the result to other neurons.

### B. Using the non-linear response of spin-torque nano-oscillators for computing

The non-linear response of spin-torque nano-oscillators combined with their stability and long life time enables neuromorphic computing. This has been experimentally demonstrated recently by our team and collaborators [3]. We have used time-multiplexing to emulate a full neural network of 400 neurons with a single spin-torque nano-oscillator (Fig. 4). The temporal connections between neurons come from the finite relaxation time of oscillator due to magnetization relaxation. Using the framework of reservoir computing [5], we have demonstrated that the multiplexed nano-oscillator could recognize spoken digits from 0 to 9 with a precision up to 99.6 %, which is as good as much larger neurons and software simulations. These results show that spin-torque nano-oscillators can be readily used as neurons.

## III. Spin-torque nano-oscillators can learn to classify the microwave signals they receive

### A. Spin-torque nano-oscillators response is highly sensitive to incoming microwave signals

Spin-torque nano-oscillators feature an oustanding tunability [6]. For instance, their frequency can be varied by more than 50 % by changing the injected direct current or the applied magnetic field. Due to this property, their dynamics can easily be influenced by weak external microwave currents or magnetic fields. In particular, spin-torque nano-oscillators have a high ability to phase lock to external microwave signals (Fig. 5) and to mutually synchronize (Fig. 6). From a neural network point of view, this means that spintronic oscillators as neurons in layer $k+1$ have the ability to adapt their response to the microwave outputs broadcasted by neurons in layer $k$. In other words, microwave inter-neuron communication is possible in spin-torque nano-oscillator networks, opening a path to ultra-fast processing. How such a network would function and compute is an open question, but many possibilities exist. Fig. 7 gives an insight on how this would work: if neuron $i$ in layer $k$ and neuron j in layer $k+1$ are synchronized, they share the same frequency, which means that the synapse $w_{ij}$ that connects them is strong. On the other hand, if neuron $i$ and $j$ have very different frequencies, the synaptic weight $w_{ij}$ is weak.

### B. Learning to classify microwave inputs

We have recently given a proof of concept of microwave signal classification through synchronization [4]. The experimental network is composed of four spin-torque nano-oscillators interconnected by millimeter-long wires (Fig. 8). The microwave emission from each oscillator propagates through this electrical loop and in turn modifies the dynamics of all the other oscillators. The oscillators are all-to-all coupled though this mechanism, which emulates synaptic connections. The frequency of the individual oscillators can be controlled by injecting different direct currents in the junctions. The oscillator microwave emissions are detected with a spectrum analyzer. As illustrated in Fig. 9, microwave inputs are injected in an antenna (a strip line) located just above the oscillators. Through the antenna, the microwave inputs generate a microwave magnetic field that modifies the dynamics of the oscillators. Depending on the relative frequency of oscillators and inputs, the frequency of the oscillator is pulled by the ensemble of input signals, or it can synchronize to one of the inputs. These emerging synchronization configurations can be used to classify inputs, if a given class of inputs always gives rise to the same synchronization configuration (for instance, the same set of synchronized oscillators). This behavior can be obtained by training the neural network through a gradual modification of the direct currents injected in the oscillators, which results in frequency tuning.

### C. Vowel recognition

We have recently shown that this neural network of four spin-torque nano-oscillators could be trained to classify seven American vowels (https://youtu.be/IHYnh0oJgOA). The frequencies characteristics of the vowels are accelerated and combined to generate two input signals in the microwave domain. The experimental recognition rate after training is 89% on the test data (84% after cross validation). This performance is comparable and even slightly better than that of a multilayer perceptron trained on the same task with a similar number of parameters. Indeed, in our scheme the coupled oscillators cooperate to decide which vowel is recognized. This is not the case in perceptrons where intra-layer neurons are not connected. This result demonstrates that spin-torque nano-oscillators dynamical properties can be finely tuned to learn. It also shows that their coupling and synchronization properties can be harnessed to classify.

To go further, we show here that our scheme can be extended to classify all twelve American vowels. For this we use a larger number of the twenty experimentally observed synchronization states and combine them to recognize a vowel. The currents injected in the oscillators, their frequency and the recognition rate during training are plotted in Fig. 10. The final map of synchronization states and corresponding vowels is shown in Fig. 11. We reach a recognition rate of 68.4 % on train and test datasets. This recognition rate can be largely increased in the future by increasing the number of oscillators in the system. Indeed, the number of synchronization regions that can be used and combined for recognition scales as $N^2$ where N is the number of oscillators [7].

## IV. CONCLUSION AND PERSPECTIVES

We have shown that spin-torque nano-oscillators can non-linearly transform inputs and broadcast the result of their computation. We have shown that they can also learn to classify the microwave signals they receive. Two main challenges remain to build large scale deep neural networks with these devices: fabricating synapses that can tune the inter-neuron microwave communication, and densely interconnecting neurons through these synapses. In future work, we will investigate how the multifunctionality of spintronic building blocks [8] and the possibility to assemble them in 3D [9] can enable solutions to these challenges.

ACKNOWLEDGMENT

This work was supported by the European Research Council ERC under Grant bioSPINspired 682955, the French National Research Agency (ANR) under Grant MEMOS ANR-14-CE26-0021 and by a public grant overseen by the ANR as part of the "Investissements d'Avenir" program (Labex NanoSaclay, reference: ANR-10-LABX-0035).

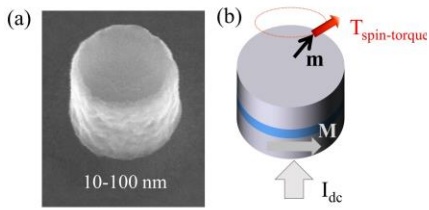

Fig. 1. (a) SEM picture of a 400 nm diameter resist dot before etching the magnetic tunnel junction stack. (b) Illustration of magnetization dynamics under spin-torque in a magnetic tunnel junction.

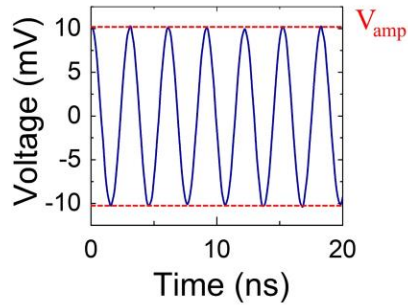

Fig. 2. Spin-torque induced voltage oscillations in a magnetic tunnel junction. The amplitude of the oscillations is indicated in red. The central stack is based on a CoFeB 2.4 nm pinned layer, MgO 1nm and FeB 6 nm free layer. The ground state is a magnetic vortex.

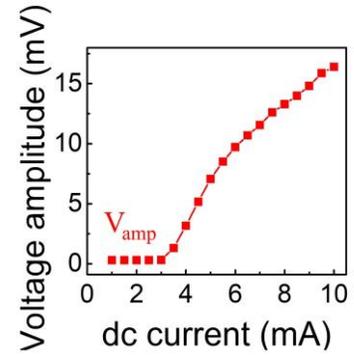

Fig. 3. Voltage amplitude as a function of the injected direct current for the junction of Fig.2.

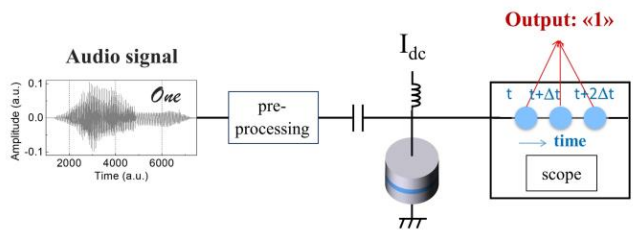

Fig. 4. Illustration for spoken digit recognition with a single time-multiplexed spin-torque nano-oscillator through reservoir computing.

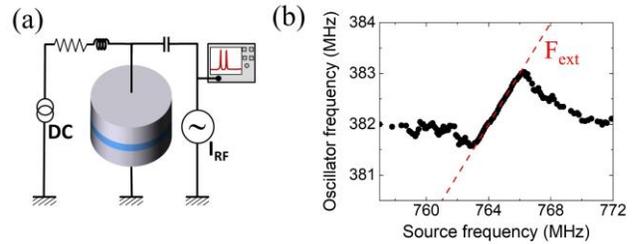

Fig. 5. (a) Schematic of the magneto-transport set-up for phase-locking (b) Oscillator frequency versus source frequency. When the oscillator is phase-locked, its frequency becomes equal to the source frequency.

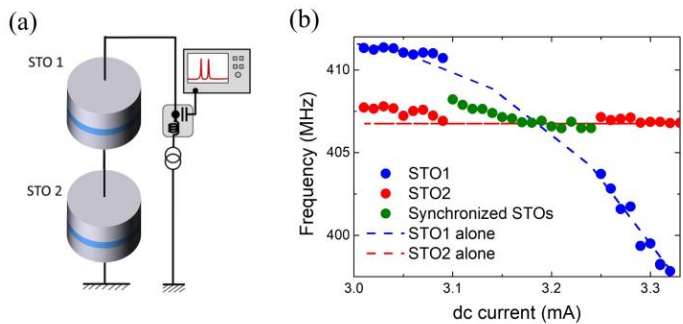

Fig. 6. a) Schematic of the magneto-transport set-up for the electrical mutual synchronization of two oscillators (b) Oscillators frequency as a function of dc current. In the locking-range, the two oscillator frequencies become identical.

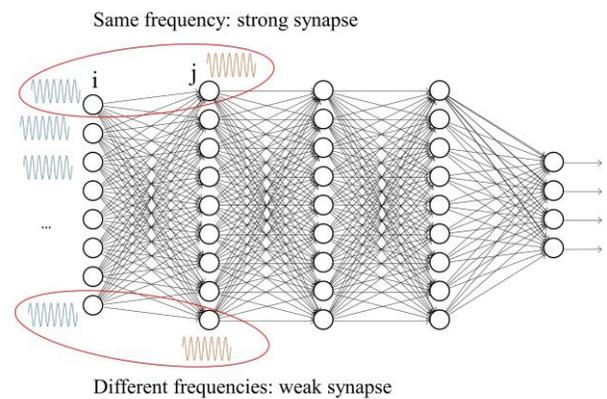

Fig. 7 Illustration of the working principle of a microwave neural network.

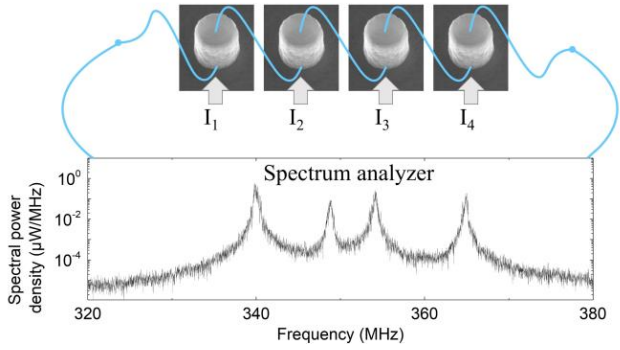

Fig. 8. Schematic of the neural network composed of four interconnected spin-torque nano-oscillators. Independent direct currents can be applied to the different oscillators, and their microwave emissions are detected with a spectrum analyzer.

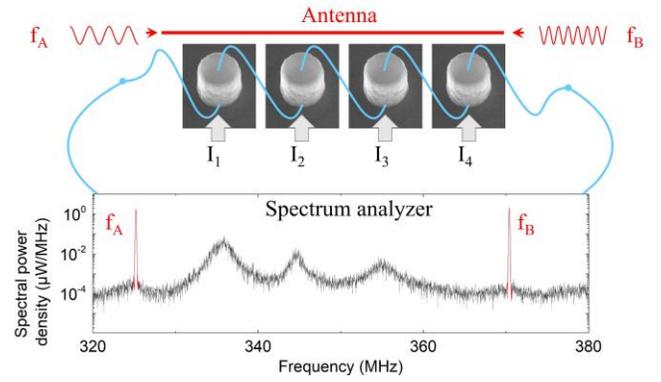

Fig. 9. Schematic of the spin-torque oscillator neural networks with microwave inputs are applied through a strip line above the oscillators. In that case, oscillator 4 is synchronized to frequency $f_B$.

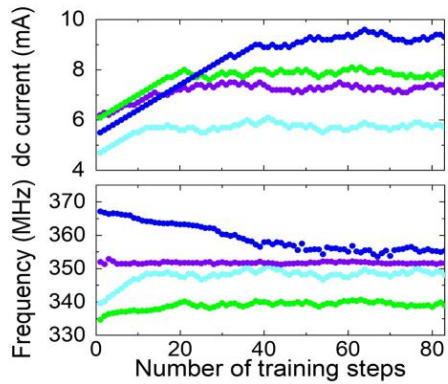
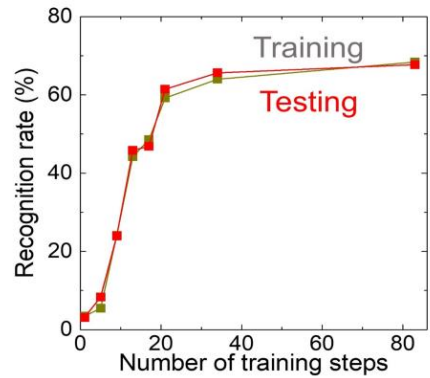

Fig. 10. Frequencies, dc currents through the oscillators and recognition rates on the train and test datasets as a function of training step.

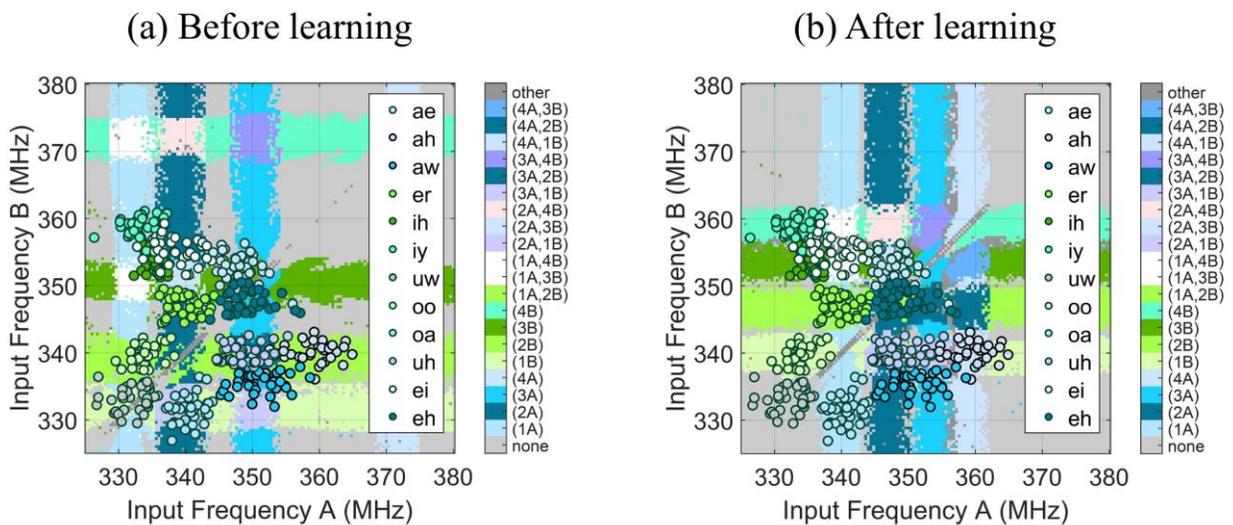

Fig. 11 Maps of synchronization regions before and after learning. The label (XA,YB) indicates that oscillator X is synchronized to source A and oscillator Y is synchronized to source B. Vowels are indicated as circles with the color of the synchronization regions in which they should be classified.